\documentclass[12pt,a4paper]{article}
\usepackage{amssymb}
\usepackage{graphicx,color}
\makeatletter
\def\rddots{\mathinner{\mkern1mu\raise\p@%
    \vbox{\kern7\p@\hbox{.}}\mkern2mu%
    \raise4\p@\hbox{.}\mkern2mu\raise7\p@\hbox{.}\mkern1mu}}
\makeatother
\makeatletter
\def\eqnarray{%
\stepcounter{equation}%
\let\@currentlabel=\theequation
\global\@eqnswtrue
\global\@eqcnt\z@
\tabskip\@centering
\let\\=\@eqncr
$$\halign to \displaywidth\bgroup\@eqnsel\hskip\@centering
$\displaystyle\tabskip\z@{##}$&\global\@eqcnt\@ne
\hfil$\displaystyle{{}##{}}$\hfil
&\global\@eqcnt\tw@$\displaystyle\tabskip\z@{##}$\hfil
\tabskip\@centering&\llap{##}\tabskip\z@\cr}
\makeatother
\newcommand{\ket}[1]{{\vert{#1}\rangle}}

\newcommand{\fukuso}{{\mathbf C}}
\newcommand{\real}{{\mathbf R}}
\newcommand{\futon}{{\bf N}}

\begin{document}

\title{\sl On the Magic Matrix by Makhlin and the B-C-H Formula in $SO(4)$}
\author{
  Kazuyuki FUJII
  \thanks{E-mail address : fujii@yokohama-cu.ac.jp }\quad and \ 
  Tatsuo SUZUKI
  \thanks{E-mail address : suzukita@gm.math.waseda.ac.jp }\\
  ${}^{*}$Department of Mathematical Sciences\\
  Yokohama City University\\
  Yokohama, 236--0027\\
  Japan\\
  ${}^{\dagger}$Department of Mathematical Sciences\\
  Waseda University\\
  Tokyo, 169--8555\\
  Japan\\
  }
\date{}
\maketitle
%
%
%
%
\begin{abstract}
  A closed expression to the Baker--Campbell--Hausdorff (B-C-H) formula 
  in $SO(4)$ is given by making use of the magic matrix by Makhlin. 
  As far as we know this is the {\bf first nontrivial example} on (semi--) 
  simple Lie groups summing up all terms in the B-C-H expansion.
\end{abstract}
%


%
%
%
%

\section{Introduction}

The Baker--Campbell--Hausdorff (B--C--H) formula is one of fundamental ones 
in elementary Linear Algebra (or Lie group). That is, we have
\[
\mbox{e}^{A}\mbox{e}^{B}=\mbox{e}^{{BCH}(A,B)}
\]
where $A$ and $B$ are elements in some algebra and
\[
{BCH}(A,B)=A+B+\frac{1}{2}[A,B]+
\frac{1}{12}\left\{[[A,B],B]+[A,[A,B]]\right\}+\cdots.
\]
See for example \cite{Vara} or \cite{YS}. 
For simplicity we call this the B--C--H expansion in the text. Although the formula is 
``elementary", it is difficult (almost impossible ?) to sum up all terms in ${BCH}(A,B)$. 
If $[A,B]$ and $A$, $B$ commute, then we have
\[
{BCH}(A,B)=A+B+\frac{1}{2}[A,B].
\]
However, this is very useful but exceptional, \cite{MS}. 

By the way, it is not difficult to give a close expression to ${BCH}(A,B)$ for the case 
of $SU(2)$ because it is easy to treat. 
We would like to use this expression in the paper.

Next, to apply the Makhlin's theorem to the problem let us explain it. 
The isomorphism 
\[
SU(2)\otimes SU(2)\cong SO(4)
\]
is one of well--known theorems in elementary representation theory and is a 
typical characteristic of four dimensional Euclidean space. 
In \cite{YMa} Makhlin gave it {\bf the adjoint expression} like
\[
F : SU(2)\otimes SU(2) \longrightarrow SO(4),\quad 
F(A\otimes B)=Q^{\dagger}(A\otimes B)Q
\]
with some unitary matrix $Q\in U(4)$. As far as we know this is the first that 
the map was given by the adjoint action. See also \cite{FOS}, where 
a bit different matrix $R$ has been used in place of $Q$ because $Q$ is of 
course not unique. 
This $Q$ ($R$ in our notation) is interesting enough and is called the magic 
matrix by Makhlin, see \cite{ZVWS} and \cite{RZ}.

Since a close expression for the B--C--H formula in the case of $SU(2)$ 
is known, we can also obtain the close expression for ${BCH}(A,B)$ in the case of 
$SO(4)$
\[
\mbox{e}^{A}\mbox{e}^{B}=\mbox{e}^{{BCH}(A,B)}\quad \mbox{for}\quad A, B \in so(4)
\]
by making use of the magic matrix by Makhlin. This is the main result in the paper.
As far as we know this is the first nontrivial example on (semi--) simple Lie groups 
summing up all terms in the B-C-H expansion.

\section{Review on the Magic Matrix}

In this section we review the result in \cite{FOS} within our necessity, which is 
a bit different from the one in \cite{YMa}.

The 1--qubit space is $\fukuso^{2}=\mbox{Vect}_{\fukuso}\{\ket{0},\ket{1}\}$ 
where
\begin{equation}
\label{eq:bra-ket}
\ket{0}=
\left(
\begin{array}{c}
 1 \\
 0
\end{array}
\right),
\quad 
\ket{1}=
\left(
\begin{array}{c}
 0 \\
 1
\end{array}
\right).
\end{equation}
Let $\{\sigma_{1}, \sigma_{2}, \sigma_{3}\}$ be the Pauli matrices acting on 
the space
\begin{equation}
\label{eq:Pauli matrices}
\sigma_{1} = 
\left(
  \begin{array}{cc}
    0 & 1 \\
    1 & 0
  \end{array}
\right), \quad 
\sigma_{2} = 
\left(
  \begin{array}{cc}
    0 & -i \\
    i & 0
  \end{array}
\right), \quad 
\sigma_{3} = 
\left(
  \begin{array}{cc}
    1 & 0 \\
    0 & -1
  \end{array}
\right).
\end{equation}

Next let us consider the 2--qubit space. 
Now we use notations on tensor product which are different from usual ones. 
That is, 
\[
\fukuso^{2}{\otimes}\fukuso^{2}=\{a\otimes b\ |\ a,b\in \fukuso^{2}\},
\]
while
\[
\fukuso^{2}\widehat{\otimes}\fukuso^{2}=
\left\{\sum_{j=1}^{k}c_{j}a_{j}\otimes b_{j}\ |\ a_{j},b_{j}\in \fukuso^{2},
\ c_{j}\in \fukuso,\ k\in \futon \right\}\cong \fukuso^{4}.
\]
Then
\[
\fukuso^{2}\widehat{\otimes}\fukuso^{2}=\mbox{Vect}_{\fukuso}
\{\ket{00},\ket{01},\ket{10},\ket{11}\}
\]
where $\ket{ab}=\ket{a}\otimes \ket{b}\ (a,b\in \{0,1\})$. 

By $H_{0}(2;\fukuso)$ we show the set of all traceless hermite matrices in 
$M(2;\fukuso)$. Then it is well--known
\[
H_{0}(2;\fukuso)=\{a\equiv a_{1}\sigma_{1}+a_{2}\sigma_{2}+a_{3}\sigma_{3}\ 
|\ a_{1},a_{2},a_{3}\in \real\}
\]
and $H_{0}(2;\fukuso)\cong su(2)$ where $su(2)$ is the Lie algebra of the 
group $SU(2)$.

\vspace{5mm}
By making use of the Bell bases 
$\{\ket{\Psi_{1}},\ket{\Psi_{2}},\ket{\Psi_{3}},\ket{\Psi_{4}}\}$ 
defined by
\begin{eqnarray}
\label{eq:Bell bases}
\ket{\Psi_{1}}&=&\frac{1}{\sqrt{2}}(\ket{00}+\ket{11}),\quad
\ket{\Psi_{2}}=\frac{1}{\sqrt{2}}(\ket{01}+\ket{10}), \nonumber \\
\ket{\Psi_{3}}&=&\frac{1}{\sqrt{2}}(\ket{01}-\ket{10}),\quad
\ket{\Psi_{4}}=\frac{1}{\sqrt{2}}(\ket{00}-\ket{11})
\end{eqnarray}
we can give the isomorphism as the adjoint action (the Makhlin's theorem) 
as follows
\[
F : SU(2)\otimes SU(2) \longrightarrow SO(4),\quad 
F(A\otimes B)=R^{\dagger}(A\otimes B)R
\]
where
\begin{equation}
R=
\left(
\ket{\Psi_{1}},-i\ket{\Psi_{2}},-\ket{\Psi_{3}},-i\ket{\Psi_{4}}
\right)
=\frac{1}{\sqrt{2}}
\left(
  \begin{array}{cccc}
    1 &  0 &  0 & -i  \\
    0 & -i & -1 &  0  \\
    0 & -i &  1 &  0  \\
    1 &  0 &  0 &  i 
  \end{array}
\right).
\end{equation}
Note that the unitary matrix $R$ is a bit different from $Q$ in \cite{YMa}.

\vspace{3mm}
Let us consider this problem in a Lie algebra level because it is in general 
not easy to treat it in a Lie group level.
\begin{center}
\input{Lie-diagram.fig}
\end{center}

\vspace{5mm} \noindent
Since the Lie algebra of $SU(2)\otimes SU(2)$ is
\[
\mathfrak{L}(SU(2)\otimes SU(2))=
\left\{i(a\otimes 1_{2}+1_{2}\otimes b)\ |\ a,b \in H_{0}(2;\fukuso)\right\},
\]
we have only to examine
\begin{equation}
\label{eq:Lie algebra level}
f(i(a\otimes 1_{2}+1_{2}\otimes b))=
iR^{\dagger}(a\otimes 1_{2}+1_{2}\otimes b)R\in 
\mathfrak{L}(SO(4))\equiv so(4).
\end{equation}
If we set $a=\sum_{j=1}^{3}a_{j}\sigma_{j}$ and $b=\sum_{j=1}^{3}b_{j}\sigma_{j}$ 
then the right hand side of (\ref{eq:Lie algebra level}) becomes
\begin{equation}
\label{eq:correspondence}
iR^{\dagger}(a\otimes 1_{2}+1_{2}\otimes b)R 
=
\left(
  \begin{array}{cccc}
    0 & a_{1}+b_{1} & a_{2}-b_{2} & a_{3}+b_{3}       \\
    -(a_{1}+b_{1}) & 0 & a_{3}-b_{3} & -(a_{2}+b_{2}) \\
    -(a_{2}-b_{2}) & -(a_{3}-b_{3}) & 0 & a_{1}-b_{1} \\
    -(a_{3}+b_{3}) & a_{2}+b_{2} & -(a_{1}-b_{1}) & 0  
  \end{array}
\right).
\end{equation}
Conversely, if
\[
A=
\left(
  \begin{array}{cccc}
    0 & f_{12} & f_{13} & f_{14}   \\
   -f_{12} & 0 & f_{23} & f_{24}   \\
   -f_{13} & -f_{23} & 0 & f_{34}  \\
   -f_{14} & -f_{24} & -f_{34} & 0  
  \end{array}
\right)\in so(4)
\]
then we obtain
\begin{equation}
RAR^{\dagger}=i(a\otimes 1_{2}+1_{2}\otimes b)
\end{equation}
with
\begin{eqnarray}
\label{eq:left-a}
&&a=a_{1}\sigma_{1}+a_{2}\sigma_{2}+a_{3}\sigma_{3}
   =\frac{f_{12}+f_{34}}{2}\sigma_{1}+
    \frac{f_{13}-f_{24}}{2}\sigma_{2}+
    \frac{f_{14}+f_{23}}{2}\sigma_{3}, \\
\label{eq:right-b}
&&b=b_{1}\sigma_{1}+b_{2}\sigma_{2}+b_{3}\sigma_{3}
   =\frac{f_{12}-f_{34}}{2}\sigma_{1}-
    \frac{f_{13}+f_{24}}{2}\sigma_{2}+
    \frac{f_{14}-f_{23}}{2}\sigma_{3}.
\end{eqnarray}
It is very interesting to note that $a$ is the self--dual part and $b$ 
the anti--self--dual one.

\vspace{3mm}
The matrix $R$ is called the magic one by Makhlin. Readers will understand 
why this is called magic through this paper.

\section{B-C-H Formula for SU(2)}

In this section we give a closed expression to the B--C--H formula for $SU(2)$, which 
is more or less well--known. See for example \cite{SW} and \cite{KE}.

First of all let us recall the well--known formula.
\begin{equation}
\mbox{e}^{i(x\sigma_{1}+y\sigma_{2}+z\sigma_{3})}
=
\cos{r}{\bf 1}+\frac{\sin{r}}{r}i(x\sigma_{1}+y\sigma_{2}+z\sigma_{3})
\end{equation}
where $r=\sqrt{x^{2}+y^{2}+z^{2}}$ and ${\bf 1}=1_{2}$ for simplicity. This is a 
simple exercise.

\vspace{3mm}
For the group $SU(2)$ it is easy to sum up all terms in  the B--C--H expansion 
by making use of the above one. Before stating the result let us prepare some 
notations. For 
\[
X=x_{1}\sigma_{1}+x_{2}\sigma_{2}+x_{3}\sigma_{3},\quad 
Y=y_{1}\sigma_{1}+y_{2}\sigma_{2}+y_{3}\sigma_{3}\in H_{0}(2,\fukuso)
\]
we set 
\[
X\ \longrightarrow\ {\bf x}=
\left(
  \begin{array}{c}
    x_{1} \\
    x_{2} \\
    x_{3}
  \end{array}
\right),\quad 
Y\ \longrightarrow\ {\bf y}=
\left(
  \begin{array}{c}
    y_{1} \\
    y_{2} \\
    y_{3}
  \end{array}
\right)
\]
and
\[
{\bf x}\cdot {\bf y}=x_{1}y_{1}+x_{2}y_{2}+x_{3}y_{3},\quad 
|{\bf x}|=\sqrt{{\bf x}\cdot {\bf x}},\quad
|{\bf y}|=\sqrt{{\bf y}\cdot {\bf y}}
\]
and
\[
-\frac{i}{2}[X,Y]\ \longrightarrow\
{\bf x}\times {\bf y}=
\left(
  \begin{array}{c}
    x_{2}y_{3}-x_{3}y_{2} \\
    x_{3}y_{1}-x_{1}y_{3} \\
    x_{1}y_{2}-x_{2}y_{1}
  \end{array}
\right).
\]

\par \vspace{5mm}
Now we are in a position to state the B--C--H formula for $SU(2)$.
\begin{equation}
\mbox{e}^{iX}\mbox{e}^{iY}=\mbox{e}^{iZ}\ :\quad 
Z=\alpha X + \beta Y + \gamma \frac{i}{2}[X,Y]
\end{equation}
where
\begin{eqnarray}
&&\alpha\equiv \alpha({\bf x},{\bf y})=
\frac{\sin^{-1}\rho}{\rho}
\frac{\sin{|{\bf x}|}\cos{|{\bf y}|}}{|{\bf x}|},\quad
\beta\equiv \beta({\bf x},{\bf y})=
\frac{\sin^{-1}\rho}{\rho}
\frac{\cos{|{\bf x}|}\sin{|{\bf y}|}}{|{\bf y}|},  \nonumber \\
&&\gamma\equiv \gamma({\bf x},{\bf y})=
\frac{\sin^{-1}\rho}{\rho}
\frac{\sin{|{\bf x}|}\sin{|{\bf y}|}}{|{\bf x}||{\bf y}|}
\end{eqnarray}
with
\begin{eqnarray*}
\rho^{2}&\equiv& \rho({\bf x},{\bf y})^{2}  \\
&=&
\sin^{2}{|{\bf x}|}\cos^{2}{|{\bf y}|}+\sin^{2}{|{\bf y}|}
-
\frac{\sin^{2}{|{\bf x}|}\sin^{2}{|{\bf y}|}}{|{\bf x}|^{2}|{\bf y}|^{2}}
\left({\bf x}\cdot {\bf y}\right)^{2}
+
\frac{2\sin{|{\bf x}|}\cos{|{\bf x}|}
\sin{|{\bf y}|}\cos{|{\bf y}|}}{|{\bf x}||{\bf y}|}
\left({\bf x}\cdot {\bf y}\right).
\end{eqnarray*}
The proof is not difficult, so is left to readers.

\par \vspace{5mm}
Some comments are in order. \\
\noindent
(1) In \cite{SW} the Euler angle's parametrisation is used. 
However, it is particular to the case of $SU(2)\cong S^{3}$ and there is no generality, 
so we don't use it in the paper.
\par \noindent 
(2) From our result it is easy to see the result in \cite{KE} by using the adjoint 
representation $Ad : SU(2)\longrightarrow SO(3)$ (see for example \cite{FOS}).

\section{B-C-H Formula for SO(4)}

In this section we also give a closed expression to the B--C--H formula for $SO(4)$ 
by use of the results in the preceding two sections. Before that let us prepare some 
notations for simplicity.

For $A,B\in so(4)$
\begin{equation}
\label{eq:two matrices}
A=
\left(
  \begin{array}{cccc}
    0 & f_{12} & f_{13} & f_{14}   \\
   -f_{12} & 0 & f_{23} & f_{24}   \\
   -f_{13} & -f_{23} & 0 & f_{34}  \\
   -f_{14} & -f_{24} & -f_{34} & 0  
  \end{array}
\right),\quad
B=
\left(
  \begin{array}{cccc}
    0 & g_{12} & g_{13} & g_{14}   \\
   -g_{12} & 0 & g_{23} & g_{24}   \\
   -g_{13} & -g_{23} & 0 & g_{34}  \\
   -g_{14} & -g_{24} & -g_{34} & 0  
  \end{array}
\right)
\end{equation}
we can set
\[
RAR^{\dagger}=i({\bf a}_{1}\otimes {\bf 1}+{\bf 1}\otimes {\bf a}_{2}),\quad
RBR^{\dagger}=i({\bf b}_{1}\otimes {\bf 1}+{\bf 1}\otimes {\bf b}_{2})
\]
and
\begin{eqnarray*}
&&{\bf a}_{1}
   =\frac{f_{12}+f_{34}}{2}\sigma_{1}+
    \frac{f_{13}-f_{24}}{2}\sigma_{2}+
    \frac{f_{14}+f_{23}}{2}\sigma_{3},\
{\bf a}_{2}
   =\frac{f_{12}-f_{34}}{2}\sigma_{1}-
    \frac{f_{13}+f_{24}}{2}\sigma_{2}+
    \frac{f_{14}-f_{23}}{2}\sigma_{3}, \\
&&{\bf b}_{1}
   =\frac{g_{12}+g_{34}}{2}\sigma_{1}+
    \frac{g_{13}-g_{24}}{2}\sigma_{2}+
    \frac{g_{14}+g_{23}}{2}\sigma_{3},\
{\bf b}_{2}
   =\frac{g_{12}-g_{34}}{2}\sigma_{1}-
    \frac{g_{13}+g_{24}}{2}\sigma_{2}+
    \frac{g_{14}-g_{23}}{2}\sigma_{3}
\end{eqnarray*}
and
\[
{\bf \vec{a}}_{1}=
\left(
  \begin{array}{c}
  \frac{f_{12}+f_{34}}{2} \\
  \frac{f_{13}-f_{24}}{2} \\
  \frac{f_{14}+f_{23}}{2}
  \end{array}
\right),
\quad
{\bf \vec{a}}_{2}=
\left(
  \begin{array}{c}
  \frac{f_{12}-f_{34}}{2} \\
 -\frac{f_{13}+f_{24}}{2} \\
  \frac{f_{14}-f_{23}}{2}
  \end{array}
\right);\quad
{\bf \vec{b}}_{1}=
\left(
  \begin{array}{c}
  \frac{g_{12}+g_{34}}{2} \\
  \frac{g_{13}-g_{24}}{2} \\
  \frac{g_{14}+g_{23}}{2}
  \end{array}
\right),
\quad
{\bf \vec{b}}_{2}=
\left(
  \begin{array}{c}
  \frac{g_{12}-g_{34}}{2} \\
 -\frac{g_{13}+g_{24}}{2} \\
  \frac{g_{14}-g_{23}}{2}
  \end{array}
\right),
\]
and
\begin{eqnarray*}
&&\alpha_{1}=\alpha({\bf \vec{a}}_{1},{\bf \vec{b}}_{1}),\quad
\beta_{1}=\beta({\bf \vec{a}}_{1},{\bf \vec{b}}_{1}),\quad
\gamma_{1}=\gamma({\bf \vec{a}}_{1},{\bf \vec{b}}_{1}),    \\
&&\alpha_{2}=\alpha({\bf \vec{a}}_{2},{\bf \vec{b}}_{2}),\quad
\beta_{2}=\beta({\bf \vec{a}}_{2},{\bf \vec{b}}_{2}),\quad
\gamma_{2}=\gamma({\bf \vec{a}}_{2},{\bf \vec{b}}_{2}).
\end{eqnarray*}
See the preceding two sections. Then we have
\begin{eqnarray}
\mbox{e}^{A}\mbox{e}^{B}
&=&
R^{\dagger}R\mbox{e}^{A}R^{\dagger}R\mbox{e}^{B}R^{\dagger}R  \nonumber \\
&=&
R^{\dagger}\mbox{e}^{RAR^{\dagger}}\mbox{e}^{RBR^{\dagger}}R  \nonumber \\
&=&
R^{\dagger}
\mbox{e}^{i({\bf a}_{1}\otimes {\bf 1}+{\bf 1}\otimes {\bf a}_{2})}
\mbox{e}^{i({\bf b}_{1}\otimes {\bf 1}+{\bf 1}\otimes {\bf b}_{2})}
R  \nonumber \\
&=&
R^{\dagger}
\left(\mbox{e}^{i{\bf a}_{1}}\otimes \mbox{e}^{i{\bf a}_{2}}\right)
\left(\mbox{e}^{i{\bf b}_{1}}\otimes \mbox{e}^{i{\bf b}_{2}}\right)
R  \nonumber \\
&=&
R^{\dagger}
\left(\mbox{e}^{i{\bf a}_{1}}\mbox{e}^{i{\bf b}_{1}}\right)\otimes
\left(\mbox{e}^{i{\bf a}_{2}}\mbox{e}^{i{\bf b}_{2}}\right)
R  \nonumber \\
&=&
R^{\dagger}
\mbox{e}^{i(\alpha_{1}{\bf a}_{1}+\beta_{1}{\bf b}_{1}+
\gamma_{1}\frac{i}{2}[{\bf a}_{1},{\bf b}_{1}])}
\otimes
\mbox{e}^{i(\alpha_{2}{\bf a}_{2}+\beta_{2}{\bf b}_{2}+
\gamma_{2}\frac{i}{2}[{\bf a}_{2},{\bf b}_{2}])}
R  \nonumber \\
&=&
R^{\dagger}
\mbox{e}^{i\left\{
(\alpha_{1}{\bf a}_{1}+\beta_{1}{\bf b}_{1}+
\gamma_{1}\frac{i}{2}[{\bf a}_{1},{\bf b}_{1}])\otimes {\bf 1}
+
{\bf 1}\otimes (\alpha_{2}{\bf a}_{2}+\beta_{2}{\bf b}_{2}+
\gamma_{2}\frac{i}{2}[{\bf a}_{2},{\bf b}_{2}])
\right\}}
R  \nonumber \\
&=&
\mbox{e}^{iR^{\dagger}\left\{
(\alpha_{1}{\bf a}_{1}+\beta_{1}{\bf b}_{1}+
\gamma_{1}\frac{i}{2}[{\bf a}_{1},{\bf b}_{1}])\otimes {\bf 1}
+
{\bf 1}\otimes (\alpha_{2}{\bf a}_{2}+\beta_{2}{\bf b}_{2}+
\gamma_{2}\frac{i}{2}[{\bf a}_{2},{\bf b}_{2}])
\right\}R} \nonumber \\
&=&
\mbox{e}^{{BCH}(A,B)}
\end{eqnarray}
where
\begin{eqnarray}
\label{eq:B-C-H closed}
{BCH}(A,B)
&=&iR^{\dagger}\left\{
\left(\alpha_{1}{\bf a}_{1}+\beta_{1}{\bf b}_{1}+
\gamma_{1}\frac{i}{2}[{\bf a}_{1},{\bf b}_{1}]\right)\otimes {\bf 1}
+
{\bf 1}\otimes \left(\alpha_{2}{\bf a}_{2}+\beta_{2}{\bf b}_{2}+
\gamma_{2}\frac{i}{2}[{\bf a}_{2},{\bf b}_{2}]\right)
\right\}R   \nonumber \\
&=&
\left(
  \begin{array}{cccc}
     0 & (12) & (13) & (14)   \\
    -(12) & 0 & (23) & (24)   \\
    -(13) & -(23) & 0 & (34)  \\
    -(14) & -(24) & -(34) & 0
  \end{array}
\right)
\end{eqnarray}
whose entries are
\begin{eqnarray*}
(12)&=&
\alpha_{1}\frac{f_{12}+f_{34}}{2}+\beta_{1}\frac{g_{12}+g_{34}}{2}-
\gamma_{1}\left(\frac{f_{13}-f_{24}}{2}\frac{g_{14}+g_{23}}{2}-
\frac{f_{14}+f_{23}}{2}\frac{g_{13}-g_{24}}{2}\right)+ \\
&{}&
\alpha_{2}\frac{f_{12}-f_{34}}{2}+\beta_{2}\frac{g_{12}-g_{34}}{2}-
\gamma_{2}\left(-\frac{f_{13}+f_{24}}{2}\frac{g_{14}-g_{23}}{2}+
\frac{f_{14}-f_{23}}{2}\frac{g_{13}+g_{24}}{2}\right), \\
(13)&=&
\alpha_{1}\frac{f_{13}-f_{24}}{2}+\beta_{1}\frac{g_{13}-g_{24}}{2}-
\gamma_{1}\left(\frac{f_{14}+f_{23}}{2}\frac{g_{12}+g_{34}}{2}-
\frac{f_{12}+f_{34}}{2}\frac{g_{14}+g_{23}}{2}\right)+ \\
&{}&
\alpha_{2}\frac{f_{13}+f_{24}}{2}+\beta_{2}\frac{g_{13}+g_{24}}{2}-
\gamma_{2}\left(-\frac{f_{14}-f_{23}}{2}\frac{g_{12}-g_{34}}{2}+
\frac{f_{12}-f_{34}}{2}\frac{g_{14}-g_{23}}{2}\right), \\
(14)&=&
\alpha_{1}\frac{f_{14}+f_{23}}{2}+\beta_{1}\frac{g_{14}+g_{23}}{2}-
\gamma_{1}\left(\frac{f_{12}+f_{34}}{2}\frac{g_{13}-g_{24}}{2}-
\frac{f_{13}-f_{24}}{2}\frac{g_{12}+g_{34}}{2}\right)+ \\
&{}&
\alpha_{2}\frac{f_{14}-f_{23}}{2}+\beta_{2}\frac{g_{14}-g_{23}}{2}-
\gamma_{2}\left(-\frac{f_{12}-f_{34}}{2}\frac{g_{13}+g_{24}}{2}+
\frac{f_{13}+f_{24}}{2}\frac{g_{12}-g_{34}}{2}\right), \\
(23)&=&
\alpha_{1}\frac{f_{14}+f_{23}}{2}+\beta_{1}\frac{g_{14}+g_{23}}{2}-
\gamma_{1}\left(\frac{f_{12}+f_{34}}{2}\frac{g_{13}-g_{24}}{2}-
\frac{f_{13}-f_{24}}{2}\frac{g_{12}+g_{34}}{2}\right) \\
&{}&
-\alpha_{2}\frac{f_{14}-f_{23}}{2}-\beta_{2}\frac{g_{14}-g_{23}}{2}+
\gamma_{2}\left(-\frac{f_{12}-f_{34}}{2}\frac{g_{13}+g_{24}}{2}+
\frac{f_{13}+f_{24}}{2}\frac{g_{12}-g_{34}}{2}\right), \\
(24)&=&
-\alpha_{1}\frac{f_{13}-f_{24}}{2}-\beta_{1}\frac{g_{13}-g_{24}}{2}+
\gamma_{1}\left(\frac{f_{14}+f_{23}}{2}\frac{g_{12}+g_{34}}{2}-
\frac{f_{12}+f_{34}}{2}\frac{g_{14}+g_{23}}{2}\right)+ \\
&{}&
\alpha_{2}\frac{f_{13}+f_{24}}{2}+\beta_{2}\frac{g_{13}+g_{24}}{2}-
\gamma_{2}\left(-\frac{f_{14}-f_{23}}{2}\frac{g_{12}-g_{34}}{2}+
\frac{f_{12}-f_{34}}{2}\frac{g_{14}-g_{23}}{2}\right), \\
(34)&=&
\alpha_{1}\frac{f_{12}+f_{34}}{2}+\beta_{1}\frac{g_{12}+g_{34}}{2}-
\gamma_{1}\left(\frac{f_{13}-f_{24}}{2}\frac{g_{14}+g_{23}}{2}-
\frac{f_{14}+f_{23}}{2}\frac{g_{13}-g_{24}}{2}\right) \\
&{}&
-\alpha_{2}\frac{f_{12}-f_{34}}{2}-\beta_{2}\frac{g_{12}-g_{34}}{2}+
\gamma_{2}\left(-\frac{f_{13}+f_{24}}{2}\frac{g_{14}-g_{23}}{2}+
\frac{f_{14}-f_{23}}{2}\frac{g_{13}+g_{24}}{2}\right).
\end{eqnarray*}

\par \vspace{5mm}
We could obtain the closed expression for the B--C--H formula and 
this is our main result in the paper. Note that we can transform (\ref{eq:B-C-H closed}) 
into various forms, which will be left to readers. 

\par \noindent
As far as we know this is the first nontrivial example summing up all terms 
in the B--C--H expansion.

\section{Discussion}

In this letter we studied the B--C--H formula for the case of $SO(4)$ and 
obtained the closed expression by making use of the formula for the case of $SU(2)$ 
and the magic matrix $R$ by Makhlin.

The Makhlin's matrix is essential in the case of $SO(4)$ and the readers should 
recognize the reason why it is called magic. It will be used in Quantum Computation 
and Mathematical Physics moreover, see for example \cite{KF3}, \cite{KF2}.

Last, we would like to make a comment on some generalization of our work. 
In \cite{SW} an interesting method to calculate the B--C--H formula for the case of 
$SU(n)$ has been presented. However, to perform it explicitly may be difficult even 
for the case $SU(4)$ \footnote{In fact, the calculation given in \cite{SW} for 
the case $SU(4)$ is incomplete} 
\[
\mbox{e}^{iX}\mbox{e}^{iY}=\mbox{e}^{iBCH(X,Y)}\quad \mbox{for}\quad 
X,Y\in H_{0}(4,\fukuso).
\]

To apply our method to the same case may be useful, which will be reported elsewhere.


\end{document}